\documentclass[prc,aps,nofootinbib,twocolumn]{revtex4} 
\usepackage{epsfig}
\usepackage{graphicx}
\usepackage{ulem}
\begin{document}
\title{Large Amplitude motion with a stochastic mean-field approach
\footnote{Contribution to the International Conference 
"Nuclear Structure and Related Topics" , July 2-July 7  (2012), Dubna, Russia.}
}
\author{Lacroix Denis}
\affiliation{GANIL, CEA and CNRS/IN2P3, Bo\^ite Postale 55027, 14076 Caen Cedex, France}
\author{ Sakir Ayik }
\affiliation{Physics Department, Tennessee Technological University, Cookeville, TN 38505, USA }
\author{ Bulent Yilmaz}
\affiliation{Physics Department, Ankara University, Tandogan 06100, Ankara, Turkey}
\author{Kouhei Washiyama}
\affiliation{RIKEN Nishina Center, Wako, Saitama 351-0198, Japan}
\begin{abstract}
In the stochastic mean-field approach, an ensemble of initial conditions is considered to incorporate correlations beyond the mean-field.
Then each starting pont is propagated separately using the Time-Dependent Hartree-Fock equation of motion. This approach provides a rather simple tool to better
describe
fluctuations compared to the standard TDHF. Several illustrations are presented showing that this theory can be rather effective to treat
the dynamics close to a quantum phase transition. Applications to fusion and transfer reactions demonstrate the great improvement 
in the description of mass dispersion.
\end{abstract} 
\maketitle

\section{Introduction}
\label{intro}

The mean-field description of a many-body system, i.e. the Hartree-Fock (HF) and/or time-dependent Hartree-Fock theory (TDHF), provides a simple tool for descriptions of certain aspects of complex quantum systems. However, it is well known that the mean-field approximation is suitable for the description of mean values of one-body observables, while quantum fluctuations of collective variables are severely underestimated. A second limitation of mean-field dynamics is that it cannot describe spontaneous symmetry breaking during dynamical evolution. If certain symmetries are present in the initial state, these symmetries are preserved during the evolution.
We have recently shown that a stochastic mean-field (SMF) approach \cite{Lac04,Ayi08} where the TDHF evolution is replaced by a set of mean-field evolution with properly chosen initial conditions. It will be shown that this approach can be a suitable tool to go beyond mean-field and describe the evolution of a system close to a quantum phase-transition \cite{Lac12}. 
In a series of article, we applied the SMF approach to describe transport properties in fusion reaction. 
Transport coefficients related to dissipation and fluctuations have been obtained \cite{Ayi09,Was09,Yil11} that are crucial to understand the physics of Heavy-Ion collisions around the Coulomb barrier. A summary of recent results is presented.

\section{The stochastic mean-field theory}

In a mean-field approach, the nuclear many-body dynamical problem is replaced 
by a system of particles interacting through a common self-consistent mean-field. 
Then, the information on the system is contained in the one-body density matrix $\rho$
that evolves according to the so-called TDHF equation:
\begin{eqnarray}
i\hbar \frac{\partial}{\partial t} \rho = [h[\rho],\rho],
\label{eq:puretdhf}
\end{eqnarray}
where 
$h[\rho]\equiv \partial {\cal E}(\rho) /{\partial \rho}$ denotes the mean-field Hamiltonian.
While quite successful in the description of some aspects of nuclear structure and reactions \cite{Sim08}, it
is known to not properly describe fluctuations of one-body degrees of freedom, i.e. correlations. 
Numerous approaches have been proposed either deterministic or 
stochastic to extended mean-field and describe fluctuations in collective space (see ref. \cite{Lac04}
and reference therein). 
Most often, these approaches are too complex to be applied in realistic situations
with actual computational power.  A second limitation of mean-field dynamics is that it can not 
describe spontaneous symmetry breaking during dynamical evolution. If certain symmetries are present 
in the initial state, these symmetries are preserved during the evolution  \cite{Bla86,Rin80}. 

The Stochastic Mean-Field (SMF) has been recently shown  to provide a suitable 
answer for the description of fluctuations as well as of the symmetry breaking process while keeping
the attractive aspects of mean-field. Let us assume that the aim is to improve 
the description of a system that, at the mean-field level and time $t_0$, is described by a density of the form:
\begin{eqnarray}
\rho (t_0)= \sum_i 
|\varphi_i (t_0) \rangle n_i \langle \varphi_i (t_0)|. \label{eq:densmf}
\end{eqnarray}
Note that, this density can describe either a pure Slater determinant ($n_i = 0,1$) or more generally 
an initial many-body density of the form:
\begin{eqnarray}
\hat D = \frac{1}{z}\exp \left(\sum \lambda_i a^\dagger_i a_i \right)  \label{eq:mbdensity}
\end{eqnarray}  
where $z$ is a normalization factor while $(a^\dagger_i , a_i)$ are the creation/annihilation operators associated 
to the canonical basis $|\varphi_i \rangle $. 
Then, the mean-field evolution, Eq. (\ref{eq:puretdhf}) reduces to the evolution of the set of single-particle states 
\begin{eqnarray}
i\hbar \frac{\partial}{\partial t} |\varphi_i (t) \rangle = h[\rho] |\varphi_i (t) \rangle,
\label{eq:phitdhf}
\end{eqnarray}
while keeping the occupation numbers constant.

In the SMF approach, a set of initial one-body densities  
\begin{eqnarray}
\rho^{\lambda} (t_0)& = & \sum_{ij}  |\varphi_i \rangle \rho^{\lambda}_{i, j} (t_0) \langle \varphi_j |  \label{eq:ini}
\end{eqnarray}
is considered, where $\lambda$ denotes a given initial density. The density matrix components
$\rho^{\lambda}_{i, j}$ are chosen in such a way that initially, the density obtained by averaging over 
different initial conditions identifies the density (\ref{eq:densmf}).

\begin{figure}
 \includegraphics[width=6.cm]{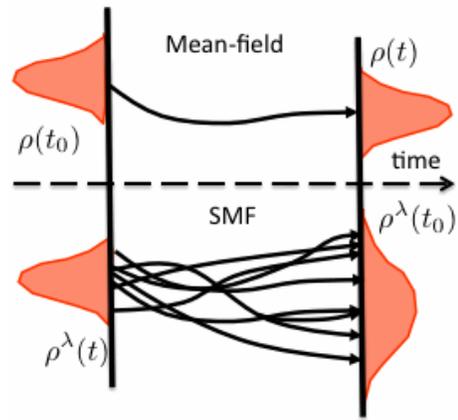}
\caption{Illustration of the quantal mean-field approach (top) where a single density is propagated and 
of the SMF approach (bottom) 
where a set of densities are chosen and where each initial density is propagated independently from the others.}
\label{fig1:lacroix}       
\end{figure}

It was shown in ref. \cite{Ayi08} that a convenient choice for the statistical properties of the initial sampling is 
\begin{eqnarray}
\rho^{\lambda}_{i, j} (t_0) = \delta_{ij} n_i + \delta \rho^{\lambda}_{i j} (t_0), \label{eq:mean}
\end{eqnarray}
where $\delta \rho^{\lambda}_{i j} (t_0)$ are mean-zero random Gaussian numbers while
\begin{eqnarray}
\overline{ \delta \rho^\lambda_{i j } (t_0)\delta \rho^{\lambda *}_{ kl }(t_0)} &=&  \frac{1}{2} 
\delta_{i  l }\delta_{j  k }\left( n^\alpha_i (1-n^\beta_j) + n^\beta_j (1 - n^\alpha_i)\right). \label{eq:fluc}
\end{eqnarray}
The average is taken here on initial conditions. In this approach, each initial condition given by 
Eq. (\ref{eq:ini}) is evolved with its own mean-field independently from the other trajectories, i.e. 
\begin{eqnarray}
i\hbar \frac{\partial}{\partial t} |\varphi^{\lambda}_i (t) \rangle = h[\rho^{\lambda}] |\varphi^{\lambda}_i (t) \rangle,
\label{eq:puresmf}
\end{eqnarray}
while keeping  the density matrix components constant.  Therefore, the evolution 
along each trajectory is similar to standard mean-field propagation and can be implement with existing codes.
A schematic illustration of the standard mean-field and stochastic mean-field is given in figure \ref{fig1:lacroix}.

\section{From densities to observables: Ehrenfest formulation of the Stochastic Mean-Field Theory}

The mean-field theory is a quantal approach and even if it usually underestimates fluctuations of collective 
observables in the nuclear physics context, these fluctuations are non-zero. Within mean-field theory, the expectation value of
an observable $\hat O$ is obtained through $\langle \hat O \rangle = {\rm Tr} (\hat O \hat D)$ where $\hat D$ has the 
form (\ref{eq:mbdensity}). Accordingly, the quantal average and fluctuations of a one-body observable $\hat Q$ along the 
mean-field trajectory are given by:
\begin{eqnarray}
\langle \hat Q \rangle = \sum_i \langle \varphi_i(t) |\hat Q | \varphi_i(t) \rangle n_i \label{eq:avermf}
\end{eqnarray}
and 
\begin{eqnarray}
\sigma^2_Q(t) =\langle \hat Q^2 \rangle - \langle \hat Q \rangle^2  = \sum_i | \langle \varphi_i(t) |\hat Q | \varphi_i(t) \rangle|^2 n_i (1-n_i). \label{eq:flucmf} 
\end{eqnarray}

An important aspect of the SMF approach is that the quantum expectation value is replaced by 
a classical statistical average over the initial conditions. Denoting by $Q^{\lambda}(t)$ the value 
of the observable at time $t$ for a given event, fluctuations are obtained using 
\begin{eqnarray}
\sigma^2_{\rm SMF}(t) & = & \overline{(Q^{\lambda}(t) - Q(t))^2} 
\end{eqnarray}   
where $Q(t) = \overline{Q^{\lambda}(t)}$. The statistical properties of initial conditions insures that 
quantal fluctuations [$\sigma^2_Q$] and statistical [$\sigma^2_{\rm SMF}$] fluctuations are equal at initial time.
Note that such a classical mapping is a known technique to simulate quantum objects and might even
be exact in some cases \cite{Her84,Kay94}.

In practice, it might be advantageous to select few collective degrees of freedom instead of the full one-body density 
matrix. At the mean-field level, the evolution of a set of one-body observable $\hat Q_i$ is given by the Ehrenfest theorem:
\begin{eqnarray}
i\hbar \frac{d}{dt} \langle \hat Q_i \rangle =  \langle [ \hat Q_i, \hat H]  \rangle
\end{eqnarray}  
If a complete set of one-body observables is taken, for instance if we consider full set of operators$ \{ a^\dagger_i a_j  \}$, one recovers 
eq. (\ref{eq:puretdhf}). In many situations, one might further reduce the evolution to a restricted set of relevant degrees of freedoms in such a way that 
the mean-field approximation leads to a closed set of equations between them, i.e.
\begin{eqnarray}
i\hbar \frac{d}{dt} \langle \hat Q_i \rangle =  {\cal F} \left( \langle \hat Q_1 \rangle, \cdots , \langle \hat Q_n \rangle \right). \label{eq:rel}
\end{eqnarray}   
Starting from this equation, one can also formulate the SMF theory directly in the selected space of degrees of freedom by considering 
a set of initial conditions $\{ Q^\lambda_i (t_0) \}_{i=1,n}$ and by using directly the evolution:
\begin{eqnarray}
i\hbar \frac{d}{dt} \hat Q^{\lambda}_i (t) =  {\cal F} \left( \hat Q^{\lambda}_1 (t), \cdots , \hat Q^{\lambda}_n (t) \right). 
\end{eqnarray}  
for each initial condition $\lambda$. Note that statistical properties, i.e. first and second moments, of initial conditions should be computed 
using the conditions (\ref{eq:mean}) and (\ref{eq:fluc}).

\section{Illustrations}

In recent years, we have applied the SMF approach either to schematic models or to realistic situations encountered 
in nuclear reactions where mean-field alone was unable to provide a suitable answer. Some examples are briefly 
discussed below.

\subsection{Many-body dynamics near a saddle point}

As mentioned in the introduction, the mean-field theory alone cannot break a symmetry 
by itself. The symmetry breaking can often be regarded as the presence of a saddle point 
in a collective space while the absence of symmetry breaking in mean-field just means that the system will stay 
at the top of the saddle if it is left here initially. Such situation is well illustrated in the Lipkin-Meshkov-Glick
model.  This model consists of $N$ particles 
distributed in two N-fold degenerated single-particle states separated by an energy $\varepsilon$. The associated Hamiltonian 
is given by (taking $\hbar=1$),
\begin{eqnarray}
H = \varepsilon J_z - V(J^2_x  -  J_y^2) , 
\label{eq:hamillipkin}
\end{eqnarray}
where $V$ denotes the interaction strength while $J_i$ ($i=x$, $y$, $z$), are the quasi-spin operators defined as
\begin{eqnarray} 
J_z &=& \frac{1}{2} \sum_{p=1}^{N} \left(c^\dagger_{+,p}c_{+,p} - c^\dagger_{-,p}c_{-,p}\right) , \nonumber \\
J_x &=& \frac{1}{2} (J_+ + J_-), ~~~J_y = \frac{1}{2i} (J_+ - J_-)
\end{eqnarray}
with $J_+ = \sum_{p=1}^{N} c^\dagger_{+,p}c_{-,p}$, $J_- = J_+^\dagger$ and where 
$c^\dagger_{+,p}$ and $c^\dagger_{-,p}$ are creation operators associated with the upper and lower single-particle levels.
In the following, energies and times are given in $\varepsilon$ and $\hbar/\varepsilon$ units respectively.

\begin{figure}[htbp] 
\begin{center}
  \includegraphics[width=8.cm]{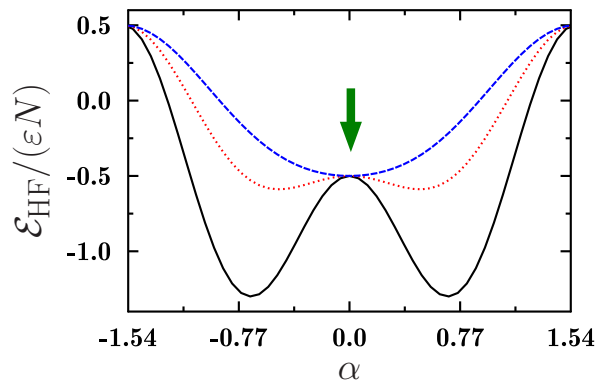}
  \end{center}
\caption{(color online). Evolution of the Hartree-Fock energy ${\cal E}_{\rm HF}$ as a function of $\alpha$ for $\chi=0.5$ (dashed line), 
$\chi=1.8$ (doted line) and $\chi=5$ (solid line) for $N=40$ particles.
The arrow indicates the initial condition used in the SMF dynamics.} 
\label{fig2:lacroix} 
\end{figure}

It could be shown that the TDHF dynamic can be recast as a set of coupled equations between 
the expectation values of the quasi-spin operators $j_i \equiv  \langle J_i \rangle/N$ (for $i=x$,  $y$ and $z$) given by:
\begin{eqnarray}
\frac{d}{dt} 
\left(
\begin{array} {c}
 j_x     \\
  j_y  \\
 j_z 
\end{array}
\right)
&=& \varepsilon 
\left(
\begin{array} {ccc}
 0   & -1 + \chi  j_z  &  \chi   j_y  \\
 1+ \chi   j_z    &  0 & \chi  j_x  \\
 -2 \chi   j_y & -2 \chi   j_x  & 0
\end{array}
\right)
\left(
\begin{array} {c}
 j_x    \\
 j_y \\
 j_z 
\end{array}
\right) \label{eq:tdhf}
\end{eqnarray}
where $\chi = V(N-1) / \varepsilon$. Note that, this equation of motion is nothing but a special case 
of eq. (\ref{eq:rel}) where the information is contained in the three quasi-spin components. 
 To illustrate the symmetry breaking in this model it is convenient 
to display the Hartree-Fock energy ${\cal E}_{\rm HF} $ as a function of the $j_z$ component (Fig. \ref{fig2:lacroix}). 
Note that, here the order parameter $\alpha = \frac{1}{2} {\rm acos}(-j_z/2)$ is used for conveniency. 
When the strength parameter is larger than  a critical value ($\chi > 1$), the parity symmetry is broken in $\alpha$ direction. 
For $\chi > 1$,  if the system is initially at the position indicated by the arrow in Fig. \ref{fig2:lacroix}, with TDHF it will remain at this point, i.e. 
this initial condition is a stationary solution of Eq.  (\ref{eq:tdhf}). 

Following the strategy discussed above, a SMF approach can be directly formulated in collective space where initial random conditions
for the spin components are taken. Starting from the statistical properties (\ref{eq:mean}) and (\ref{eq:fluc}), it could be shown that the
quasi-spins should be  initially  sampled according to Gaussian probabilities with 
first moments given by \cite{Lac12}:
\begin{eqnarray}
\overline{j^\lambda_x(t_0)}=\overline{j^\lambda_y(t_0)} = 0,
\end{eqnarray}
and second moments determined by, 
\begin{eqnarray}
\overline{j^\lambda_x(t_0)j^\lambda_x(t_0)} = \overline{j^\lambda_y(t_0) j^\lambda_y(t_0)} = \frac{1}{4N}. 
\end{eqnarray}
while the $z$ component is a non fluctuating quantity.
\begin{figure}[htbp] 
\begin{center}
  \includegraphics[width=8.cm]{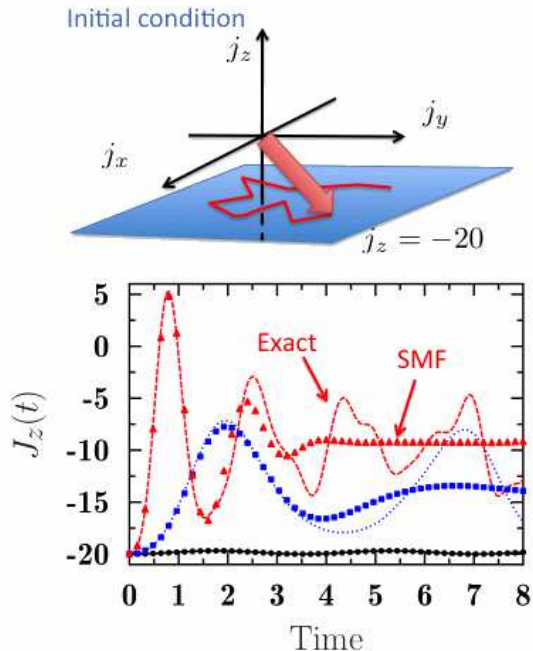}
  \end{center}
\caption{(color online) Top: illustration of the initial sampling used for the SMF theory in the collective space of quasi-spins.
Bottom:  Exact evolution of the $z$ quasi-spin component obtained when the initial state is $|j,-j\rangle$ for three 
different values of $\chi$: $\chi = 0.5$ (solid line), $\chi=1.8$ (dotted line) and $\chi=5.0$ (dashed line) for $N=40$ particles. The corresponding 
results obtained with the SMF simulations are shown with circles, squares and triangles respectively. (adapted from \cite{Lac12})} 
\label{fig3:lacroix} 
\end{figure} 
  
\begin{figure}[htbp] 
\begin{center}
  \includegraphics[width=8.cm]{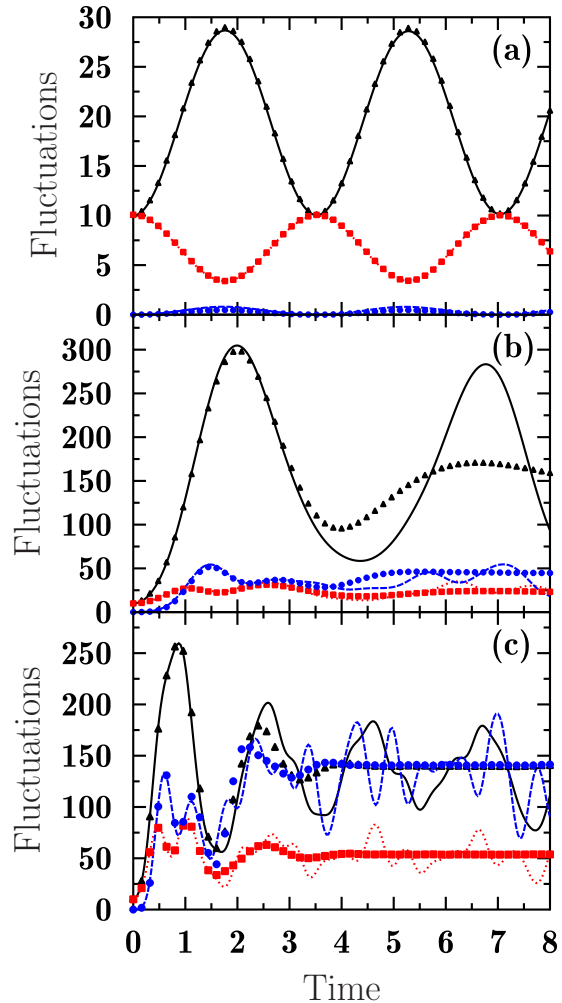}
  \end{center}
\caption{(color online)  Exact evolution of dispersions of quasi-spin operators obtained when the initial state is $|j,-j\rangle$ for three 
different values of $\chi$, from top to bottom $\chi = 0.5$ (a), $\chi=1.8$ (b) and $\chi=5.0$ (c) are shown. In each case, 
solid, dashed and dotted lines correspond to $\sigma^2_x (t)$, $\sigma^2_y (t)$ and $\Delta^2_z (t)$, respectively.
In each case, results of the SMF simulations are shown with triangles ($\sigma^2_x$), squares ($\sigma^2_y$) and 
circles ($\sigma^2_z$). (taken from \cite{Lac12})} 
\label{fig4:lacroix} 
\end{figure}

An illustration of the initial sampling (top) and of  results obtained by 	
averaging mean-field trajectories with different initial conditions is shown in Fig. \ref{fig3:lacroix} 
and compared to the exact dynamic.
As we can see from the figure, while the original mean-field gives constant quasi-spins as a function of time, the SMF approach greatly improves 
the dynamics and follows the exact evolution up to a certain time that depends on the interaction strength. As shown in Fig. \ref{fig4:lacroix}, the stochastic approach not only improves the description of the mean-value of one-body observables but also the fluctuations. 

\section{Application to nuclear reactions}

\begin{figure}[htbp] 
\begin{center}
  \includegraphics[width=8.cm]{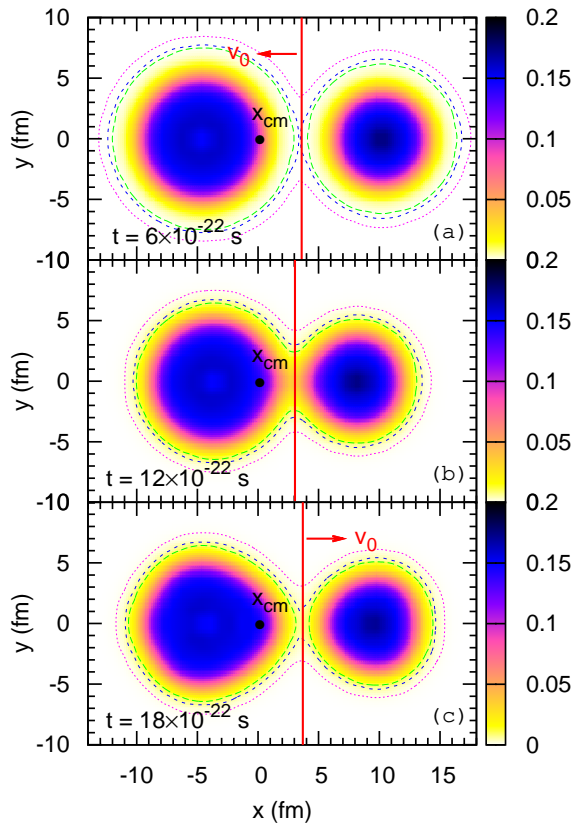} 
  \end{center}
\caption{(color online). Snapshots of the nucleon density profiles on the reaction plane, $\rho(x,y,z=0)$, are indicated by contour plots for the central collision
of $^{40}$Ca + $^{90}$Zr system at $E_{\rm cm}=97$ MeV in units of fm$^{-3}$. The black dot is the center of mass point. The red lines indicate the positions of the window $x_0$ and $v_0=dx_0/dt$ denotes velocity of the window. (taken from \cite{Yil11})} 
\label{fig5:lacroix} 
\end{figure}

The SMF has been recently used to deduce transport coefficients associated to momentum dissipation 
or mass transfer during reactions from a fully microscopic theory \cite{Ayi09,Was09,Yil11}. The TDHF theory provides 
a powerful way to get insight nuclear reaction and treat various effects like deformation, nucleon transfer, fusion, ... 
in a quantal transport theory. An illustration of nuclear densities obtained at various time of the $^{40}$Ca + $^{90}$Zr
reactions is given in Fig. \ref{fig5:lacroix}.

The mean-field approach does include the so-called one-body dissipation associated to the deformation of the system and/or to the exchange of
particles. For instance, considering a set of observables, denoted generically $\mathbf{Q} \equiv \{\hat Q_i \}$, 
like the relative distance, relative momentum, angular momentum between nuclei 
or the number of nucleons inside one of the nucleus, 
it is possible to reduce the TDHF evolution and obtain classical equations of motion of the form:
\begin{eqnarray}
\frac{\partial Q_i}{\partial t} & = & F(\mathbf{Q},t) - \sum_{j} \nu_{ij}(\mathbf{Q},t) Q_j  ,
\end{eqnarray} 
where $F$  is an eventual driving force while $\nu$ corresponds to drift coefficients. For instance, the nucleus-nucleus 
interaction potential and 
energy loss associated to internal dissipation has been extracted in ref. \cite{Was08,Was09} using such formula. 

When TDHF is extended to incorporate initial fluctuations, the equation of motion itself becomes a stochastic process:
\begin{eqnarray}
\frac{\partial Q^\lambda_i}{\partial t} & = & F(\mathbf{Q^\lambda},t) - \sum_{j} \nu_{ij}(\mathbf{Q^\lambda},t) Q^\lambda_j  + \delta Q^\lambda_i .
\end{eqnarray} 
For short time, the average drifts $\bar \nu_{ij}$ should identify with the TDHF one while the extra term is a random variables that leads to dispersion around
the mean trajectory. In the Markov limit, one can define the diffusion coefficient 
\begin{eqnarray}
\overline{\delta Q^\lambda_i (t) \delta Q^\lambda_j (t)} & = & 2 \delta(t-t') D_{ij}(t) .
\end{eqnarray} 
 
This mapping has been recently used to not only study dissipative process but also estimate 
fluctuations properties in the momentum and mass exchange. Denoting by $D_{AA}(t)$ the diffusion coefficient associated 
with mass, fluctuations in mass of the target and/or projectile can be computed using the simple formula  
\begin{eqnarray}
\sigma^2_{AA}(t) \simeq 2 \int_0^t D_{AA}(s)ds.
\label{eq:sigma}
\end{eqnarray}

\begin{figure}[htbp] 
\begin{center}
  \includegraphics[width=8.cm]{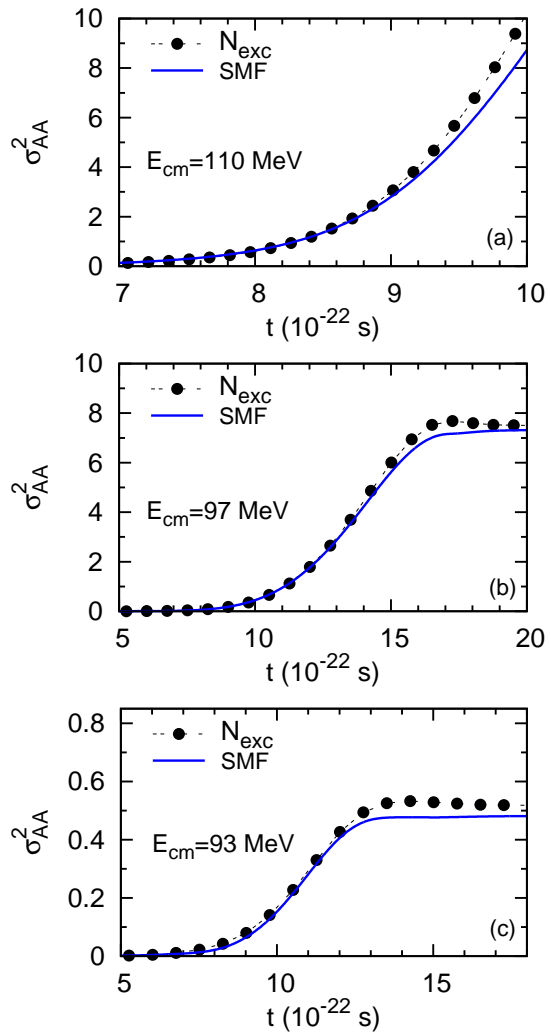}
  \end{center}
\caption{ Variances of fragment mass distributions are plotted versus time in collisions of  $^{40}$Ca + $^{90}$Zr system at three different center-of-mass energies. The dotted lines denote total number of exchanged nucleons until a given time $t$. (taken from \cite{Yil11})} 
\label{fig6:lacroix} 
\end{figure}

In figure \ref{fig6:lacroix}, an example 
of estimated variances during the asymmetric reaction $^{40}$Ca + $^{90}$Zr  is shown as a function of time 
in the case of fusion reaction (top)
or below the Coulomb barrier (middle and bottom panel).
All cases correspond to central collisions. Note that below the Coulomb barrier 
the target and projectile re-separate after having exchanged few nucleons corresponding to transfer reactions.
In general, it is observed that the fluctuations are greatly increased compared to the original 
TDHF and are compatible with the net number of exchanged nucleons from one nucleus to the other 
(dotted line in Fig. \ref{fig6:lacroix}).

\section{Summary}

In this contribution, illustrations of the application of the stochastic mean-field theory 
are discussed. It is shown, that the introduction of initial fluctuations followed by a set of independent 
mean-field trajectories greatly improves the original mean-field picture. In particular, it seems that this approach is a powerful
to increase the fluctuations that are generally strongly underestimated in TDHF or to describe the many-body dynamics
close to a saddle point.

{\bf Acknowledgments} 
S.A., B.Y., and K.W. gratefully acknowledge GANIL for the support and warm hospitality extended to them during their visits. This work is supported in part by the US DOE Grant No. DE-FG05-89ER40530.

\end{document}